\documentclass{sfb676}

\begin{document}
\title{Top-Quark Physics at the LHC}

\author{{\slshape Sven-Olaf Moch$^1$, J\"urgen Reuter$^2$}\\[1ex]
$^1$II. Institut f\"ur Theoretische Physik, Universit\"at Hamburg, Germany\\
$^2$DESY, Hamburg, Germany
}

\contribID{B11}

\desyproc{PUBDB-2018-00782}
\acronym{SFB 676 -- Particles, Strings and the Early Universe} 
\doi  

\maketitle

\begin{abstract}
%
%
We report on the precision determination of the top-quark mass 
to next-to-next-to-leading order in QCD in well-defined renormalization
schemes using data from the Large Hadron Collider for single-top and top-quark pair production.
We also discuss the calibration of the so-called Monte Carlo top-quark mass parameter 
which is determined from a comparison to events with top-quark decay products. 
The implications of the measured value of the top-quark mass for conclusions 
about the stability of the electroweak vacuum state of our Universe are illustrated.
At future lepton colliders, we provide for the first time matched
exclusive calculations valid both at the top threshold and in the
continuum, also fully differentially. In addition, we calculate fully
off-shell top-pair production (also with an associated Higgs boson) at
next-to-leading order in QCD, which allows to extract the top-Yukawa
coupling with an unprecedented precision. 

%
%
\end{abstract}

\section{Introduction}
Top-quark production features among the processes with the largest cross sections at the Large Hadron Collider (LHC). 
So far, the LHC has collected high quality data at collision energies up to $\sqrt s = 13$~TeV 
to be confronted with the high precision theory predictions of the Standard Model (SM) of particle physics. 
For the latter, the current state-of-art is based on the Quantum Chromodynamics (QCD) 
corrections up to the next-to-next-to-leading order (NNLO) 
in the strong coupling constant $\alpha_s$, 
which can be used to determine the top-quark mass $m_t$ at this theoretical accuracy.
After the confirmation of the Higgs boson's existence, 
these measured values of the SM parameters $m_t$ and $\alpha_s$ play a decisive
role in addressing questions currently in focus of the scientific community
and related to the Higgs boson's implications 
for our Universe at very early times and the stability of the electroweak vacuum.
Altogether this makes top-quark physics at the moment one of the most interesting areas in particle physics
and precision predictions for processes at present and future colliders are a necessary prerequisites.

In this article we begin with a brief review of the present theory predictions for the production 
of single top-quarks and top-quark pairs at the LHC.
Then we report on the precision determination of the top-quark mass $m_t(\mu_r)$ 
to NNLO in QCD in the $\overline{\mathrm{MS}}\,$ scheme at a chosen renormalization scale $\mu_r$ 
from measured cross sections at various center-of-mass energies at the LHC.
Related, we also discuss the procedure for the calibration of the Monte Carlo top-quark mass parameter 
used in the kinematic reconstruction of events with top-quark decay products. 
Finally, we apply the measured values of $m_t$ and $\alpha_s$ together with their present uncertainties 
to the running of the Higgs boson's self-coupling at large scales, 
which indicates stability of the ground state of the electroweak theory.

At future electron-positron colliders, such as the planned International Linear Collider (ILC), 
the top-quarks can be copiously produced by operating at center-of-mass energies
around the threshold to top-quark pair production.
With the large statistics collected under such conditions and the
significantly reduced uncertainties due to experimental systematics in
collisions with electrons and positrons, the ILC
is ideally suited for an improved precision measurement of $m_t$
as well as for investigating possible deviations from SM predictions in the
top-quark related processes. These aspects are detailed in the final section
of this article.

\section{Top-quark physics at the LHC}

\subsection{Inclusive top-quark hadro-production}

The theoretical description of both top-quark pair production 
and single-top production has reached a very high level of accuracy.
According to the standard QCD factorization the cross section is given as 
\begin{eqnarray}
\label{eq:QCDfactorization}
  \displaystyle
  \sigma_{pp \to X}(s,m_t^2) = 
  \sum\limits_{ij}\, 
  f_{i}(\mu_f^2) \otimes\,
  f_{j}(\mu_f^2) \otimes\, 
  \hat{\sigma}_{ij \to X} \left(\alpha_s(\mu_r),s,m_t^2,\mu_f^2 \right)\, 
  \, ,
\end{eqnarray}
where $\mu_f$ and $\mu_r$ are the factorization and renormalization scale 
and $s$ is the center-of-mass energy.
The parton distribution functions (PDFs) in the proton are denoted by $f_{i}$ ($i=q,{\bar q},g$)
and $\hat{\sigma}_{ij \to X}$ is the (hard) subprocess cross section for parton types
$i$ and $j$ to produce the final state $X$, which can be a single top-quark or a $t\bar t$ pair.
Both, PDFs and the partonic cross sections in Eq.~(\ref{eq:QCDfactorization}) are subject to the 
standard convolution in the parton momentum fractions, denoted as ``$\otimes$''. 
The top-quark mass in Eq.~(\ref{eq:QCDfactorization}) requires the choice of a 
renormalization scheme, which we take to be the $\overline{\mathrm{MS}}\,$
scheme, that is using $m_t(\mu_r)$.

\begin{table}[t]
\begin{center}
\renewcommand{\arraystretch}{1.3}
\begin{tabular}{|l|l|l|l|}
\hline
  & $\sqrt{s}=7$~TeV
  & $\sqrt{s}=8$~TeV
  & $\sqrt{s}=13$~TeV
\\[0.5ex]
\hline
$\sigma_{pp \to t{\bar t}}$~[pb]
  & $ 171.8 ^{\,+\,0.1}_{\,-\,5.3}\, \pm 3.4 $  
  & $ 247.5 ^{\,+\,0.0}_{\,-\,7.5}\, \pm 4.6 $  
  & $ 831.4 ^{\,+\,\phantom{0}0.0}_{\,-\,23.1}\, \pm 14.5 $  
\\[0.5ex]
\hline
\end{tabular}
\caption{
  Cross section for top-quark pair production at NNLO in QCD 
  with errors shown as $\sigma + \Delta \sigma_{\rm scale} + \Delta
  \sigma_{\rm PDF}$ at various center-of-mass energies of the LHC 
  using $m_t(m_t)=160.9$~GeV in the $\overline{\mathrm{MS}}\,$
  scheme and the ABMP16 PDF set~\cite{Alekhin:2017kpj}.
  The scale uncertainty $ \Delta \sigma_{\rm scale}$ is based on the shifts 
  for the choices $\mu_r=\mu_f=m_t(m_t)/2$ and $\mu_r=\mu_f=2m_t(m_t)$ and 
  $\Delta \sigma_{\rm PDF}$ is the 1$\sigma$ combined PDF+$\alpha_s$+$m_t$ error.
}
  \label{tab:ttbar}
\end{center}
\end{table}

The hadro-production of $t\bar t$ pairs is a QCD process at Born level, 
i.e., the leading order contributions to $\hat{\sigma}_{ij \to t\bar t}$ are proportional to $\alpha_s^2$
and the gluon-gluon-fusion process $gg \to t\bar t$ driven by the gluon PDF dominates.
Higher order corrections to the respective inclusive partonic cross sections 
have been calculated up to the NNLO in perturbative QCD~\cite{Baernreuther:2012ws,Czakon:2012zr,Czakon:2012pz,Czakon:2013goa}
and the result shows good apparent convergence of the perturbative expansion  
and greatly reduced sensitivity with respect to variations of the scales $\mu_r$ and $\mu_f$.
The latter is conventionally taken as an estimate of the residual theoretical uncertainty.

In Tab.~\ref{tab:ttbar} we quote results for the inclusive $t{\bar t}$ cross section 
at NNLO accuracy in QCD using the {\tt Hathor} code~\cite{Aliev:2010zk} with
the PDF set ABMP16~\cite{Alekhin:2017kpj} for various center-of-mass energies of the LHC, 
$\sqrt{s}=7, 8,$ and 13~TeV.
The top-quark mass taken in the $\overline{\mathrm{MS}}\,$ scheme is set to 
$m_t(m_t)=160.9$~GeV and the strong coupling to $\alpha_s^{(n_f=5)}(M_Z)=0.1147$.
The PDF uncertainties quoted represent the combined symmetric 1$\sigma$ uncertainty 
$\Delta \sigma({\rm PDF}+\alpha_s+m_t)$ arising from the variation 
of the PDF parameters, $\alpha_s$ and $m_t$ in 29 PDF sets of ABMP16.
Thanks to the high precision LHC data used in the ABMP16 fit and NNLO accuracy
in QCD the overall cross section uncertainty is significantly reduced.

%
%

\begin{figure}[p]
\centering
\includegraphics[width=0.875\textwidth]{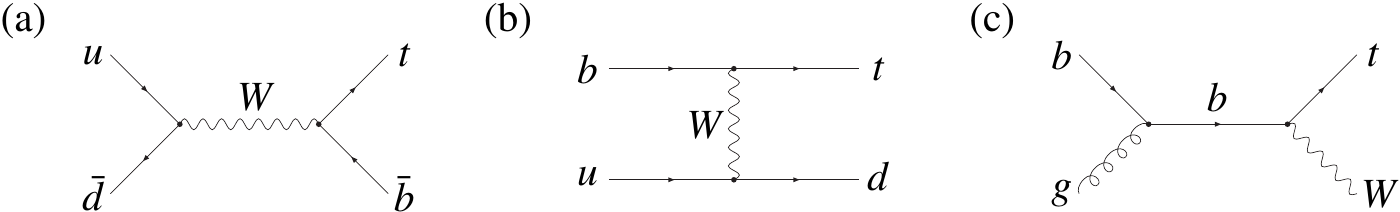}
\caption{
  Representative leading order Feynman diagrams for single top-quark production:
  (a) \mbox{$s$-channel}; (b) \mbox{$t$-channel};  (c) in association with a $W$ boson.
  Figures taken from Ref.~\cite{Alekhin:2016jjz}.
}
\label{fig:single_top_channels_diagrams}

\vspace*{10mm}

\centering
\includegraphics[width=0.48\textwidth]{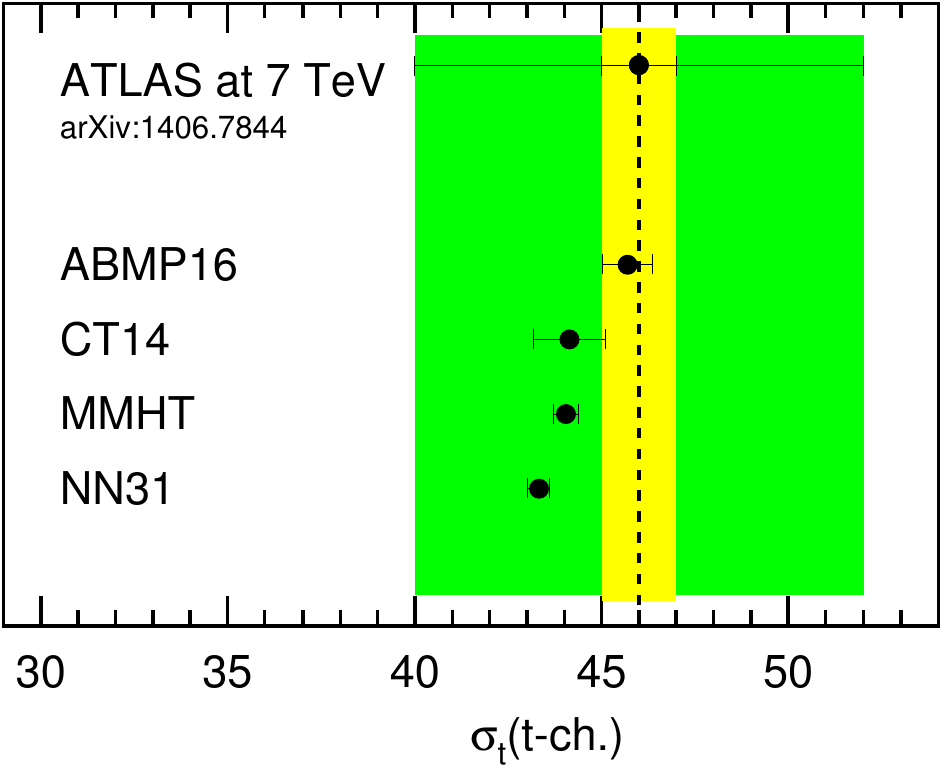}
\hfill
\includegraphics[width=0.48\textwidth]{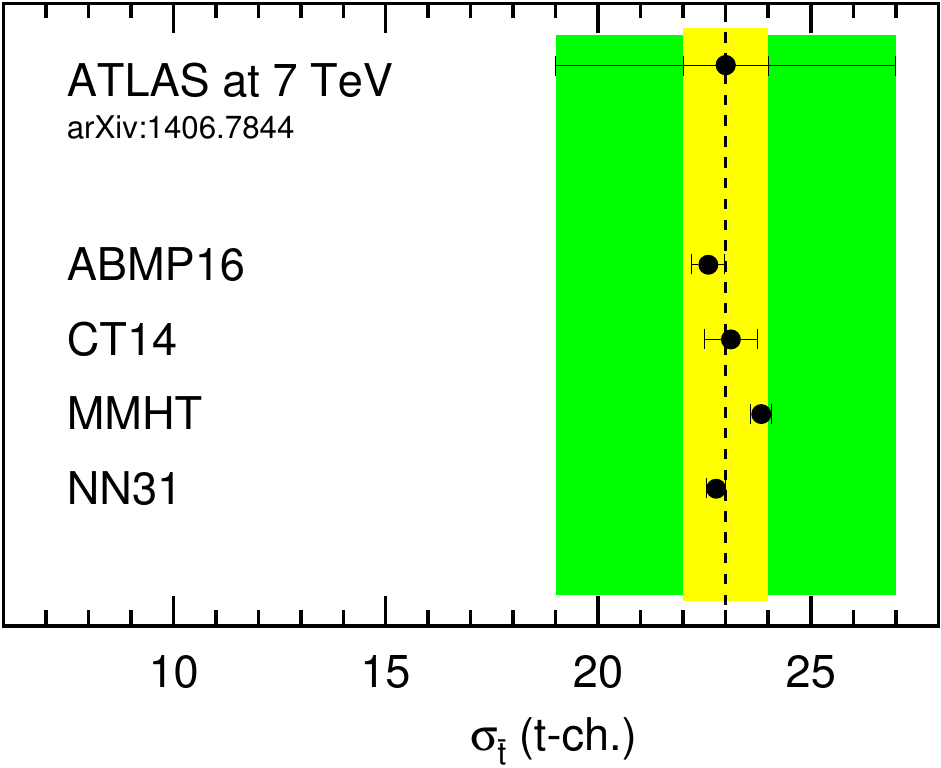}
\caption{
    Cross sections together with their $1\sigma$ PDF uncertainties
    for the $t$-channel production of single (anti)top-quarks in $pp$ collision  
    at $\sqrt s = 7$~TeV in comparison to ATLAS data~\cite{Aad:2014fwa}
    for a $\overline{\mathrm{MS}}\,$ mass $m_t(m_t) = 160.9$~GeV at the scale
    $\mu_r=\mu_f=m_t(m_t)$ with PDF sets are taken at NNLO.
    The inner (yellow) band denotes the statistical uncertainty 
    and the outer (green) band the combined uncertainty due to statistics and systematics.}
    \label{fig:sigt7}

\vspace*{10mm}

\centering
\includegraphics[width=0.48\textwidth]{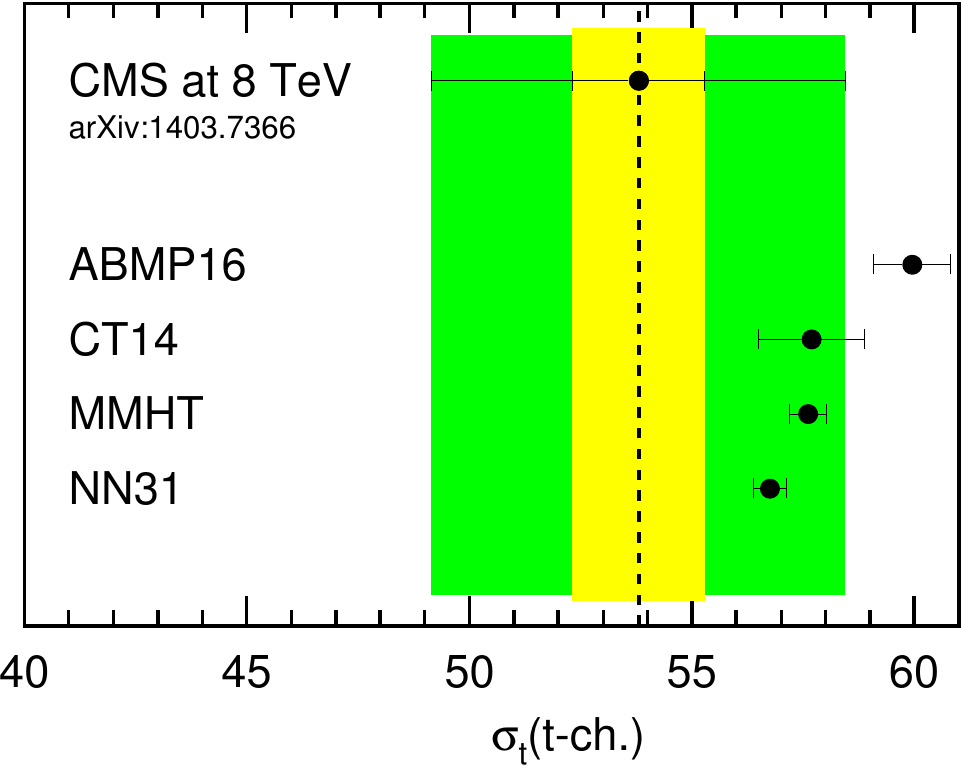}
\hfill
\includegraphics[width=0.48\textwidth]{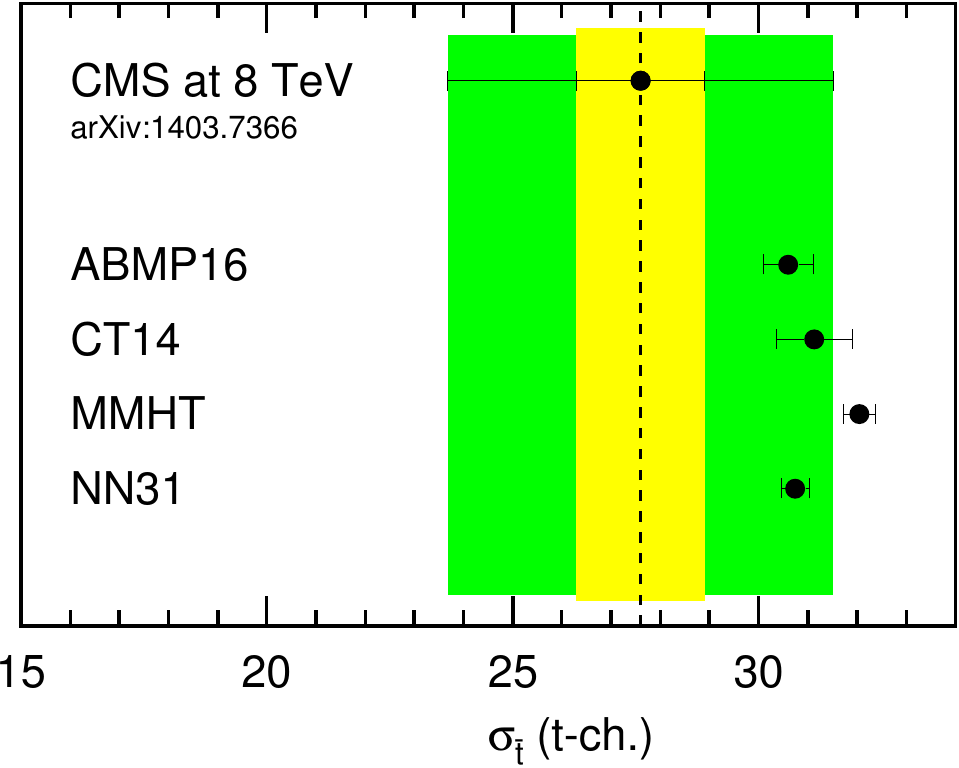}
\caption{
    Same as Fig.~\ref{fig:sigt7} for $pp$ collision  
    at $\sqrt s = 8$~TeV in comparison to CMS data~\cite{Khachatryan:2014iya}.
  }
    \label{fig:sigt8}
\end{figure}

Single-top production on the other hand generates the top-quark 
in an electroweak interaction at Born level. 
This proceeds predominantly in a vertex with a $W$-boson, top- and bottom-quark 
and its orientation assigns single-top production diagrams to different
channels.
At the LHC, the \mbox{$t$-channel} process (Fig.~\ref{fig:single_top_channels_diagrams}b) dominates, 
while the cross section for $Wt$-production (Fig.~\ref{fig:single_top_channels_diagrams}c) is typically smaller 
by one order of magnitude and the \mbox{$s$-channel} contribution
(Fig.~\ref{fig:single_top_channels_diagrams}a) is negligible.

Higher order QCD corrections to the inclusive parton cross sections $\hat{\sigma}_{ij \to t}$
for the single-top production in the \mbox{$t$-channel} are known to NNLO 
in the so-called structure function approximation~\cite{Brucherseifer:2014ama} 
(see also Ref.~\cite{Berger:2016oht}), which separately accounts 
for the QCD corrections to the light- and heavy-quark lines,
see Fig.~\ref{fig:single_top_channels_diagrams}b, 
neglecting dynamical cross-talk between the two quark lines, 
which is expected to be small due to color suppression.
Moreover, the known NNLO QCD corrections to $\hat{\sigma}_{ij \to t}$ are quite small. 
For the inclusive cross section Ref.~\cite{Brucherseifer:2014ama} has 
reported the ratios ${\sigma}_{pp \to t}^{\rm NNLO}/\sigma_{pp \to t}^{\rm NLO} \simeq -1.6\%$ 
and $\sigma_{pp \to \bar t}^{\rm NNLO}/\sigma_{pp \to\bar t}^{\rm NLO} \simeq -1.3\%$ 
at $\sqrt s = 8$~TeV for a pole mass $m_t^{\rm pole} = 173.2$ GeV.

The cross sections for \mbox{$t$-channel} single-top production are directly
proportional to the light quark PDFs. 
These are nowadays well constrained by data 
on the measured charged lepton asymmetries from $W^\pm$ gauge-boson production
at the LHC~\cite{Alekhin:2015cza}.
Therefore data on cross sections for \mbox{$t$-channel} production of single top-quarks 
offers the interesting possibility for a determination of $m_t$, 
which is not subject to strong correlations between $m_t$, $\alpha_s$ and the
gluon PDF as is the case for the cross section for $t{\bar t}$-pair production.

In Figs.~\ref{fig:sigt7} and \ref{fig:sigt8} we display 
the cross sections $\sigma_{pp \to \bar t}$ and $\sigma_{pp \to \bar t}$ at
next-to-leading order (NLO) in QCD (computed with the {\tt Hathor} library~\cite{Aliev:2010zk,Kant:2014oha})
for LHC energies of $\sqrt s = 7$ and $8$~TeV, respectively,  
using $m_t(m_t) = 160.9$ GeV with scale choice $\mu_r=\mu_f=m_t(m_t)$
and compare with data from ATLAS~\cite{Aad:2014fwa} and CMS~\cite{Khachatryan:2014iya}.
We use the NNLO PDF sets of 
ABMP16~\cite{Alekhin:2017kpj}.
CT14~\cite{Dulat:2015mca},
MMHT14~\cite{Harland-Lang:2014zoa},
and NNPDF3.1~\cite{Ball:2017nwa} 
and central value for each PDF set is complemented by the symmetrized PDF uncertainty.
Within the current large experimental uncertainties all predictions in
Figs.~\ref{fig:sigt7} and \ref{fig:sigt8} agree with the data.

%
%

\begin{figure}[t]
\centering
\includegraphics[width=0.48\textwidth]{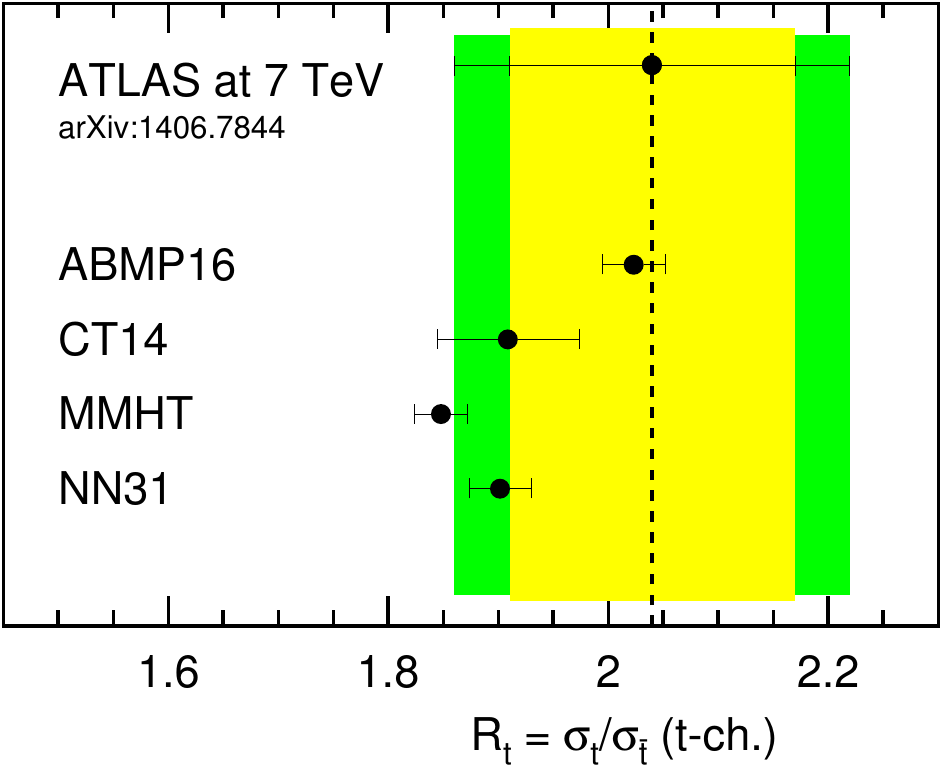}
\hfill
\includegraphics[width=0.48\textwidth]{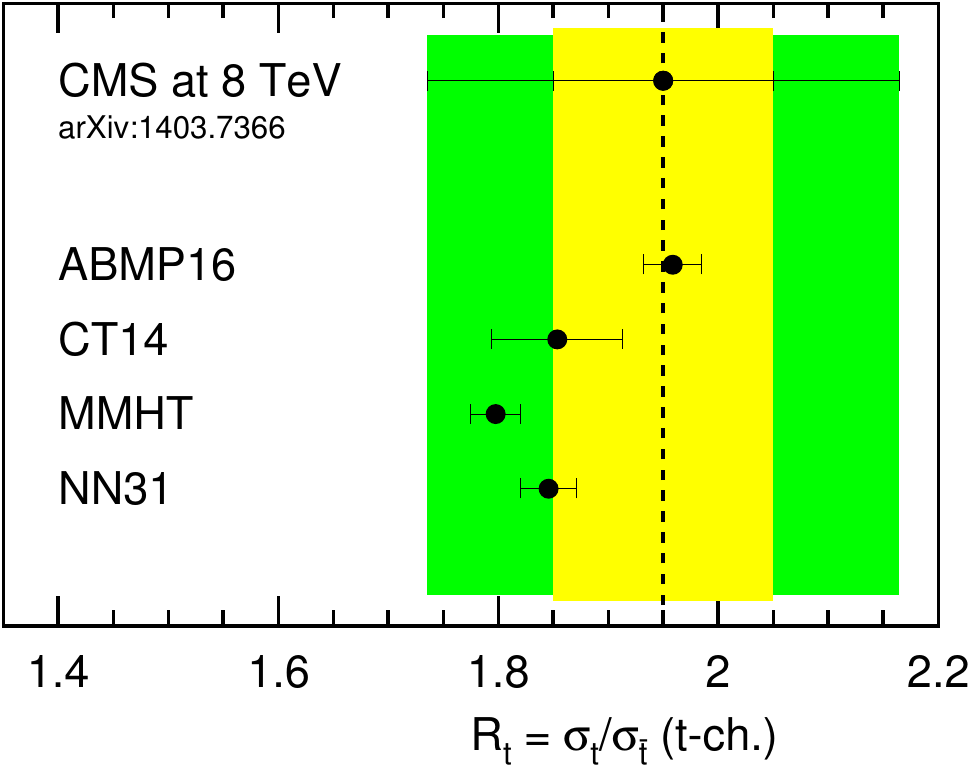}
\caption{
    Same as in Figs.~\ref{fig:sigt7} and~\ref{fig:sigt8} for the 
    ratio of cross sections $R_t=\sigma_{pp \to \bar t}/\sigma_{pp \to \bar t}$ 
    in comparison to ATLAS data~\cite{Aad:2014fwa} at $\sqrt s = 7$~TeV (left) 
    and to CMS data~\cite{Khachatryan:2014iya} at $\sqrt s = 8$~TeV (right).
  }
    \label{fig:ratio7+8}
\end{figure}

In the ratio of the cross sections $R_t=\sigma_{pp \to \bar t}/\sigma_{pp \to \bar t}$ 
on the other hand many theoretical and experimental uncertainties cancel
as shown in Fig.~\ref{fig:ratio7+8}.
This quantity is thus a very sensitive probe for the ratio of $d/u$ 
quark PDFs at large $x$.
While still all predictions for $R_t$ with the various PDF sets agree with the 
data there are systematic shifts visible so that with improved statistics 
for single-top production at $\sqrt s = 13$~TeV, this reaction might serve
as a standard candle process in the future, cf. Ref.~\cite{Alekhin:2015cza}.

%
%
%

\subsection{Top-quark mass determination}

The available high precision theory predictions for the inclusive top-quark
hadro-production can be used to extract the top-quark mass
when confronted with accurate measurements of those cross sections at the LHC. 
See, e.g., Ref.~\cite{Alekhin:2017kpj} for a recent compilation of the
respective LHC data sets and Ref.~\cite{Moch:2015fra} for a review of earlier
work on precise heavy-quark mass determinations, including charm and bottom. 

To that end, it is important to keep in mind that quark masses are not
physical observables, so that the determination of $m_t$ relies 
on comparing the parametric dependence of the theory prediction $\sigma_{\rm th}(m_t)$ 
with the experimentally measured cross section value $\sigma_{\rm exp}$.
The accuracy of the extracted top-quark mass is intrinsically limited 
by the sensitivity ${\cal S}$ of $\sigma_{\rm th}(m_t)$ to $m_t$, 
\begin{equation}
\label{eq:sensitivity}
  \left| \frac{\Delta \sigma}{\sigma} \right| 
  \, = \, {\cal S}  \times  \left| \frac{\Delta m_t}{m_t} \right|\, .
\end{equation}
For the processes under consideration the sensitivity ${\cal S}\simeq 5$ for
$t{\bar t}$ hadro-production and ${\cal S}\simeq 1.5$ for single-top 
production in the $t$-channel~\cite{Kant:2014oha}.

Next, the theory computation for $\sigma_{\rm th}(m_t)$ is performed at a given
order in perturbation theory and requires the choice of a renormalization scheme for $m_t$. 
The most common choice are the on-shell scheme with the pole mass $m_t^{\rm pole}$ 
and the $\overline{\mathrm{MS}}\,$ scheme with the running top-quark mass $m_t(\mu_r)$, 
which are then extracted at a given order in perturbation theory.
The different mass definition can be related to each other in perturbation
theory, see, e.g., Ref.~\cite{Marquard:2016dcn} and Ref.~\cite{Moch:2014tta} for a review of 
relations between these and other renormalization schemes such as the
so-called MSR mass.

\begin{figure}[t]
  \centering
  \includegraphics[width=0.55\textwidth]{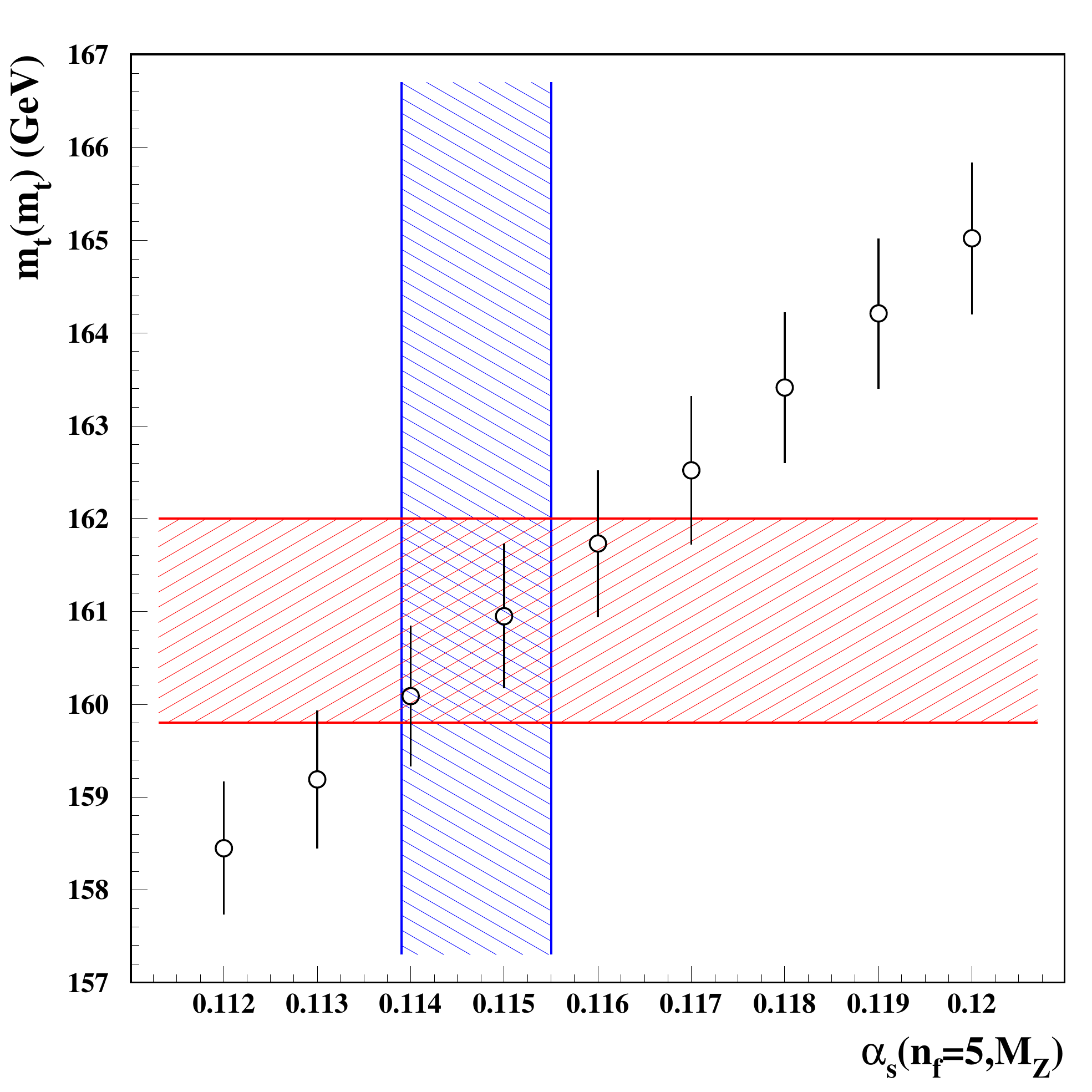}
  \caption{
    The $\overline{\mathrm{MS}}\, $ value of $m_t(m_t)$ obtained in the variants of 
    the ABMP16 analysis with the value of $\alpha_s^{(n_f=5)}(M_Z)$ fixed in comparison 
    with the $1\sigma$ bands for $m_t(m_t)$ and $\alpha_s^{(n_f=5)}(M_Z)$ 
    obtained in the nominal ABMP16 fit (left-tilted and right-tilted hatch, respectively). 
    Reprinted figure with permission from Ref.~\cite{Alekhin:2017kpj}. Copyright (2017) by the
American Physical Society.
  }
  \label{fig:scanalp}
\end{figure}

Finally, it is important to address the correlation of the $m_t$ dependence 
in $\sigma_{\rm th}(m_t)$ with all other input parameters, most 
prominently the PDFs in the proton and the strong coupling constant $\alpha_s$.
To account for those correlations any extraction of $m_t$ from
hadro-production cross sections should be performed as a global fit 
of $m_t$, the PDFs and $\alpha_s$ simultaneously~\cite{Alekhin:2017kpj}.
Shifts in $m_t$ for a given choice of a fixed value of $\alpha_s$ in 
variants of the global fit performed in Ref.~\cite{Alekhin:2017kpj} 
are illustrated in Fig.~\ref{fig:scanalp}. 
As can be seen, these are sizable and easily exceed the experimental
uncertainty on $m_t$ indicated by the vertical bars.

Using the theory predictions with a running top-quark mass $m_t(\mu_r)$ 
at NNLO for $t{\bar t}$ production~\cite{Langenfeld:2009wd} 
and for single-top production~\cite{Alekhin:2016jjz} the global analysis 
of Ref.~\cite{Alekhin:2017kpj} determines n the $\overline{\mathrm{MS}}\,$ scheme at NNLO the value
\begin{eqnarray}
\label{eq:mt}
m_t(m_t) &=& 160.9 \pm 1.1~\ensuremath{\,\mathrm{GeV}} \, .
\end{eqnarray}

The top-quark mass in Eq.~(\ref{eq:mt}) has been extracted in a well-defined renormalization scheme 
with direct relation to the parameter in the QCD Lagrangian.
A commonly used experimental procedure, on the other hand, performs a fit of
the top-quark mass parameter used in Monte Carlo simulations 
of events with top-quark decays, when comparing those simulations to 
the kinematic measurement of the top-quark decay products.
The mass determined in this way is therefore 
often referred to as the so-called top-quark Monte Carlo mass and the calibration of this mass parameter
has been a long-standing problem, since no renormalization schemes has been specified in those Monte Carlo event generators.

To overcome this problem, Ref.~\cite{Kieseler:2015jzh} has proposed a method 
to establish experimentally the relation between the top-quark mass 
$\ensuremath{m_t^{\text{MC}}}$ as implemented in a given Monte-Carlo generator and the Lagrangian mass parameter $m_t$ 
in a theoretically well-defined renormalization scheme.
The method proceeds through a simultaneous fit of
$\ensuremath{m_t^{\text{MC}}}$ and an observable with sensitivity to $m_t$, 
such as the inclusive $t{\bar t}$ cross section $\sigma_{pp \to t\bar t}$ discussed above.
In particular, this approach does not rely on any prior assumptions about the
relation between $m_t$ and $\ensuremath{m_t^{\text{MC}}}$, since 
the measured observable, e.g., $\sigma_{pp \to t\bar t}$ 
is independent of $\ensuremath{m_t^{\text{MC}}}$ and can be used subsequently for a determination of $m_t$. 

\begin{figure}[t]
\centering
\includegraphics[width=0.55\textwidth]{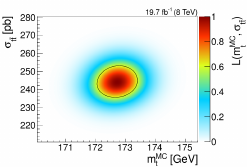}
\caption{
  Likelihood $L$ for the measured Monte Carlo mass $\ensuremath{m_t^{\text{MC}}}$ and 
  the $t\bar t$ production cross section $\ensuremath{\sigma}_{pp \to t\bar{t}}$ at a center-of-mass energy of
  $\sqrt{s} = 8$~TeV with the 1$\sigma$ uncertainty denoted by the black contour. 
	Reprinted figure with permission from Ref.~\cite{Kieseler:2015jzh}. Copyright (2016) by the
American Physical Society.
  }
\label{fig:simfit}
\end{figure}

In Fig.\ref{fig:simfit} the analysis strategy has been illustrated 
with a two parameter likelihood fit for the measured Monte Carlo mass 
$\ensuremath{m_t^{\text{MC}}}$ and the inclusive $t\bar t$ production cross section
$\ensuremath{\sigma}_{pp \to t\bar{t}}$. 
Using the measured value for $\ensuremath{\sigma}_{pp \to t\bar{t}}$ it is the
possible to extract, e.g., the pole mass $m_t^{\rm pole}$ in a subsequent step
and to determine the off-set $\Delta_m = m_t^{\rm pole}-\ensuremath{m_t^{\text{MC}}}$.
Examples from inclusive and differential cross sections for hadro-production of top-quarks
considered in Ref.~\cite{Kieseler:2015jzh} 
have led to $\Delta_m\simeq{\cal O}(2)$~GeV, depending on details of the analysis, of course.

\subsection{Stability of the electroweak vacuum}

In the light of the recent high-precision measurements of the Higgs boson mass
$m_H \,=\, 125.09 \pm 0.24$~GeV~\cite{Aad:2015zhl}, 
the measured values for the strong coupling $\alpha_s$ and the top-quark mass
$m_t$, i.e., the Yukawa coupling of the top-quark to the Higgs boson 
give rise to an intriguing coincidence.
The scalar potential $V(\phi)$ of the Higgs boson field $\phi$, given by 
\begin{equation}
  \label{eq:potential}
  V(\phi) = -\frac{m_\phi^2}{2}\phi^\dagger\phi + \lambda(\phi^\dagger\phi)^2\, ,
\end{equation}
is controlled by the scalar self-coupling $\lambda(\mu_r)$ and the 
quadratic Higgs mass parameter proportional to $m_\phi$ with the normalization 
$m_\phi=m_H$ at tree level, which combine 
in the expectation value of the electroweak vacuum as $v(\mu_r)=\sqrt{m_\phi^2(\mu_r)/(2\lambda(\mu_r))}$.
It is possible for the Higgs potential to develop a second minimum 
at field values as large as the Planck scale $M_{Pl} \simeq 10^{19}$~GeV 
in addition to $v=246$~GeV in which we live.

Investigations of the stability of the electroweak vacuum are therefore
important to answer the question, if the SM can be extended to very high
scales, where unification with gravity is expected.
Requiring $\lambda(\mu_r) \ge 0$ at all scales up to the Planck scale $M_{Pl}$ 
allows to formulate the condition for the vacuum stability 
as a lower bound on the mass of the Higgs boson as follows~\cite{Buttazzo:2013uya}
\begin{eqnarray}
\label{eq:higgsbound-1}
{\lefteqn{
m_H \,=\, 
129.6~\ensuremath{\,\mathrm{GeV}}}}\\
& &\hspace*{5mm}
\nonumber
+ 1.8 \times \left( \frac{m_t^{\rm pole} - 173.34~\ensuremath{\,\mathrm{GeV}}}{0.9} \right) 
- 0.5 \times \left( \frac{ \alpha_s^{(n_f=5)}(M_Z) - 0.1184}{0.0007} \right) \ensuremath{\,\mathrm{GeV}}
\pm 0.3~\ensuremath{\,\mathrm{GeV}}
\, ,
\end{eqnarray}
where $m_t$ and $\alpha_s$ are to be taken in the on-shell and $\overline{\mathrm{MS}}\, $ schemes,
respectively, and the uncertainty of $\pm 0.3 \ensuremath{\,\mathrm{GeV}}$ 
appears due to missing higher-order corrections (see also Ref.~\cite{Bednyakov:2015sca}).

With the values determined in Ref.~\cite{Alekhin:2017kpj} at NNLO in QCD 
for the strong coupling,\linebreak $\alpha_s^{(n_f=5)}(M_Z)=0.1147\pm 0.0008$
and the running top-quark mass $m_t(m_t)$, cf.\ Eq.~(\ref{eq:mt}),
we are in a position to update previous work~\cite{Alekhin:2012py}.
The value of $m_t$ in Eq.~(\ref{eq:mt}) 
correspond to the pole mass $m_t^{\rm pole} = 170.4 \pm 1.2$~GeV at NNLO 
so that Eq.~(\ref{eq:higgsbound-1}) leads to the bound
\begin{equation} 
\label{eq:higgsbound-2}
m_H \,=\, 126.3 \pm 2.5~\ensuremath{\,\mathrm{GeV}}
\, ,
\end{equation} 
where all uncertainties have been added in quadrature. 
Thus, within its 1$\sigma$ uncertainty 
this lower bound is compatible with the measured value $m_H \,=\, 125.09 \pm 0.24$~GeV~\cite{Aad:2015zhl} 
for the Higgs boson mass, allowing for stability up to the scale $M_{Pl}$.

\begin{figure}[t]
\centering
\includegraphics[width=0.625\textwidth]{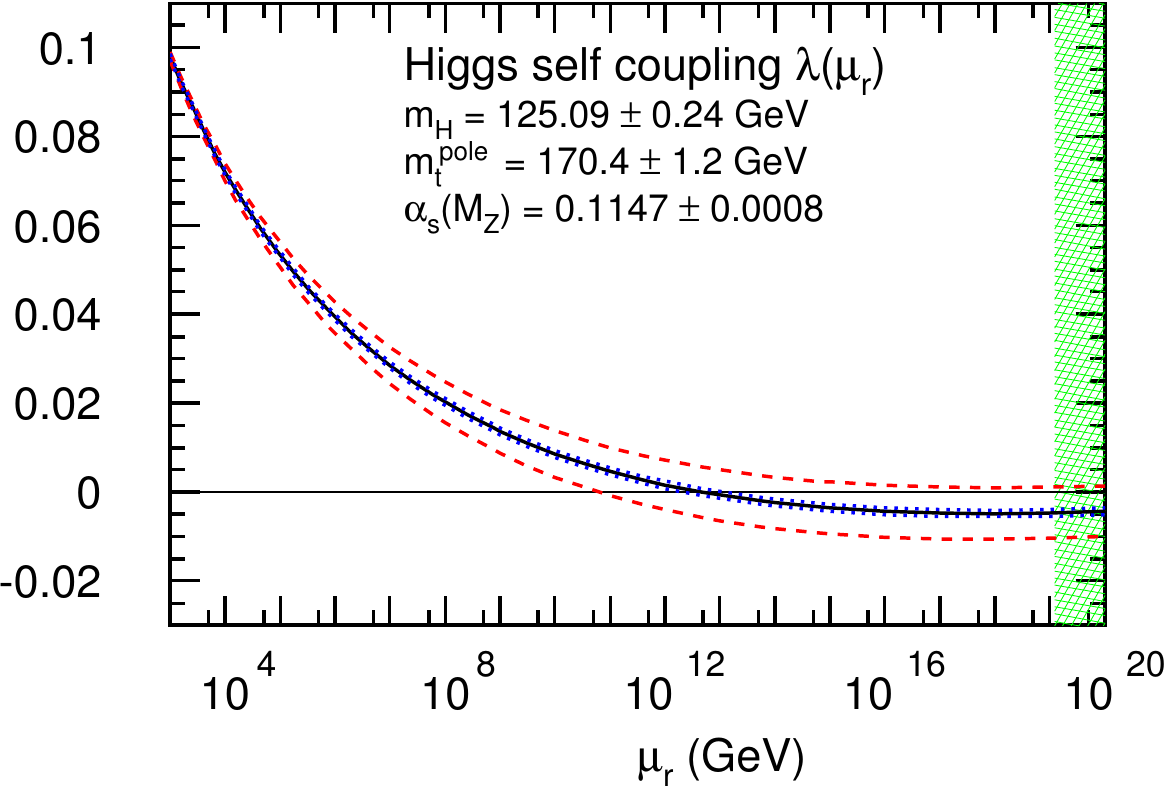}
  \caption{
    The renormalization group evolution of the Higgs boson self-coupling $\lambda$ as a function of scale $\mu_r$. 
    The dashed (red) lines denote the combined $1\sigma$ uncertainty for
    $\alpha_s^{(n_f=5)}(M_Z)$ and $m_t^{\rm pole}$ and  
    the dotted (blue) lines  the $1\sigma$ uncertainty in the value of $m_H \,=\, 125.09 \pm 0.24$.
    The range of scales $\mu_r \ge M_{Pl}$ is indicated by the hatched (green)
    band on the right.
		Reprinted figure with permission from Ref.~\cite{Alekhin:2017kpj}. Copyright (2017) by the
American Physical Society.
  }
\label{fig:lambda-rge}
\end{figure}

In a complementary way this is illustrated
in Fig.~\ref{fig:lambda-rge} showing the running of the Higgs boson self-coupling
$\lambda(\mu_r)$  in full three-loop accuracy and with
 $\alpha_s$ and $m_t$ obtained in Ref.~\cite{Alekhin:2017kpj} as the input parameters.
The computation has been performed with the code {\tt mr}, which implements matching and running of the
SM parameters~\cite{Kniehl:2016enc}.
Clearly, a vanishing Higgs self-coupling $\lambda=0$ at $M_{Pl}$ remains a scenario
which is compatible with the current values of $\alpha_s$, $m_t$ and $m_H$
within their 1$\sigma$ uncertainties. 
In addition, as follows from our analysis the value of 
$\lambda(\mu_r)$ remains strictly positive up to scales 
$\mu_r \sim {\cal O}(10^{12}\ensuremath{\,\mathrm{GeV}})$, so that no new physics needs to be invoked
in order to stabilize the electroweak vacuum.

\subsection{Beyond the SM top-quark physics}

Most models beyond the Standard Model treat the top-quark special as
it is the heaviest known particle and couples most strongly to the
electroweak symmetry breaking sector. We just discuss here one
particular aspect of new physics in the top-sector, namely the search
for so-called heavy top-partners at the LHC and the possible
measurements of their properties. These particles are predicted in
many models to compensate the large radiative corrections to the Higgs
potential from the SM top-quark. We focus here on fermionic top
partners (denoted here as $T$), where searches in specific decay channels of these particles
into the SM top-quark and a $Z$-boson, i.e., the decay $T \to tZ$, 
have been studied~\cite{Reuter:2014iya}.
This channel has been studied for hadronically decaying top-quarks by means of
so-called boosted top tagging techniques. As there is a huge mass gap expected between the
top-partner and the SM top-quark, the SM top-quark gets a considerable
boost which allows it to be treated with jet substructure techniques. Due
to the leptonically decaying $Z$ boson, this channel allows to fully
reconstruct the mass of the top-partner particle in case of a
discovery.

This has been studied in the context of simplified models, where only
the top-partner and its electroweak and QCD couplings to SM particles
are considered. Typical scenarios for such top-partners are models with composite
Higgs boson or Little Higgs models, which have also been studied
specifically in this context, cf.~\cite{Reuter:2012sd,Reuter:2013iya,Dercks:2018hgz}.

\begin{figure}[t]
\centering
\includegraphics[width=0.75\textwidth]{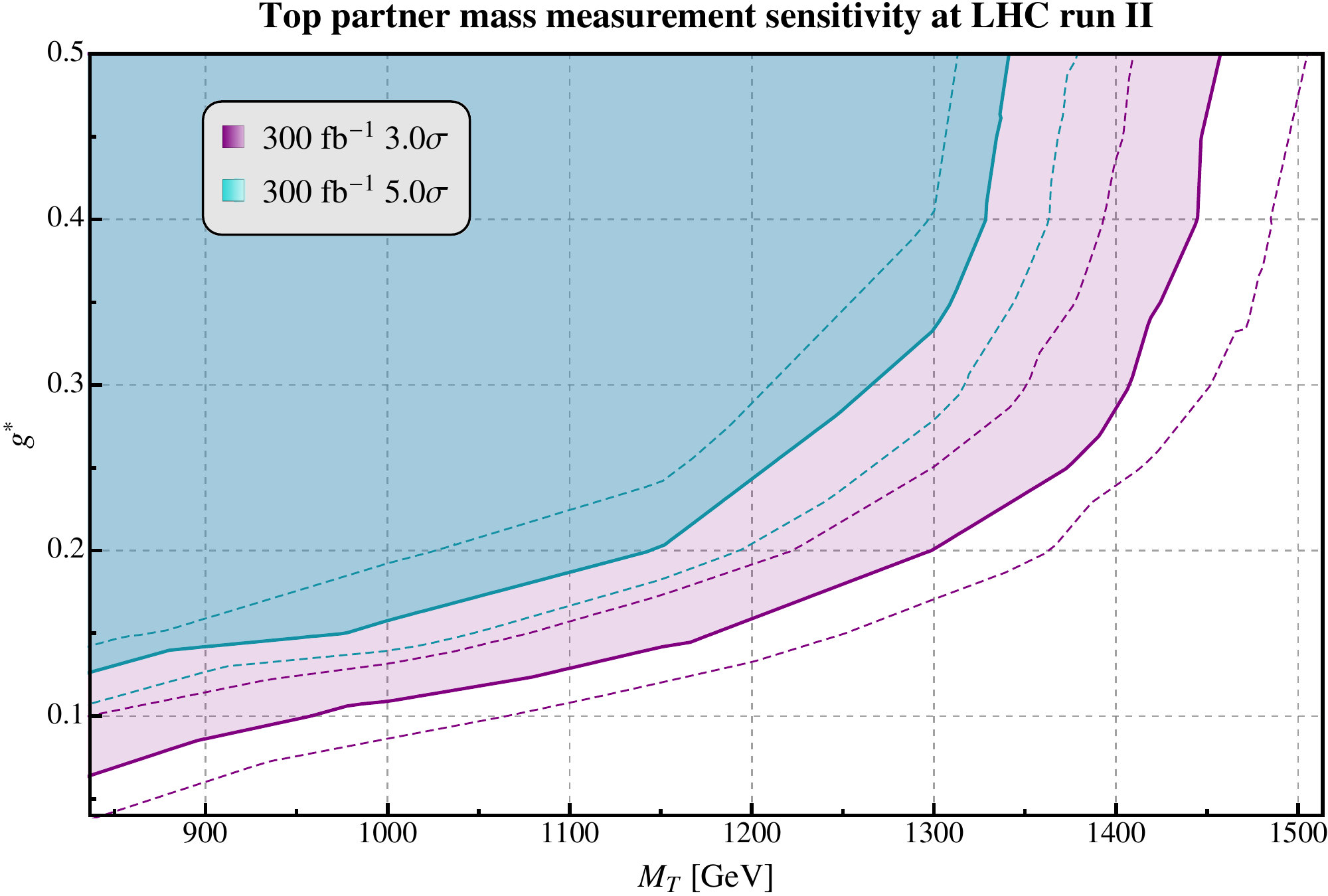}
  \caption{Sensitivity to top-partners at LHC Run 2 with $\sqrt{s} =
    13 \, \textrm{TeV}$ and $300 \, \textrm{fb}^{-1}$ of integrated
    luminosity using the $T \to tZ$ channel with hadronic boosted top
    reconstruction. Figure taken from Ref.~\cite{Reuter:2014iya}.
  }
\label{fig:toppartner_reach}
\end{figure}

Possible discoveries of a top-partner have been quantified as a function
of its mass and its universal electroweak coupling $g^\ast$. 
The sensitivity curves for 3$\sigma$ evidence as well as for
5$\sigma$ discovery are shown in Fig.~\ref{fig:toppartner_reach}. 
The shown bands include a possible additional non-statistical uncertainty
of 30\% on the visible cross section of the involved processes,
e.g., from the experimental systematics in detection efficiencies. 
If a signal peak is observed, a measurement of the top-partner's 
invariant mass is possible with a resolution 
of $\Delta m_T = \pm \, 75$ GeV around the peak, at worst.

Another project focusing on new physics in the top-quark sector
studied single-top production at the LHC as a means to measure
possible anomalous charged-current contact interactions in the
top-quark sector~\cite{Bach:2014zca}. This relies on an
effective-field theory approach which parameterizes the effect of new
physics in a model-independent way in the form of higher-dimensional
four-fermion operators for $tbff'$ contact interactions whose Wilson coefficients
can be interpreted as anomalous top-quark couplings. Here, binned
likelihood distributions over specific kinematic observables have been
studied to gain sensitivity on the Wilson coefficients. As a result it
was found that for new physics scale $\Lambda_{\rm np}$ 
of $\Lambda_{\rm np}=3$~TeV the LHC with $\sqrt{s}=14$~TeV 
and 100 fb${}^{-1}$ integrated luminosity reaches a sensitivity of the
order $\mathcal{O}(0.01-1)$ for the Wilson coefficients. Angular
distributions can serve as spin analyzers for the top-quark and allow
to resolve ambiguities in the parameter determination. 

\section{Top-quark physics at the ILC and CLIC}

Top-quark physics represents together with Higgs boson precision measurements and
the search for new physics one of the three cornerstones of the
physics program of any future electron-positron collider. The ILC
baseline design~\cite{Baer:2013cma,Fujii:2015jha} contains runs at
center-of-mass energies of both $\sqrt{s}=350$~GeV and 500 GeV, respectively,
while the baseline design of the Compact Linear Collider (CLIC)~\cite{Linssen:2012hp,Lebrun:2012hj} contains stages at $\sqrt{s}=380$~GeV,
1.4~TeV and 3~TeV, respectively. 
The new staging scenario foresees a long run of the ILC at $\sqrt{s}=250$~GeV before an energy
upgrade~\cite{Barklow:2015tja} containing the top-quark physics program.
The work carried out for this top-quark physics program within this
project encompasses fixed-order NLO QCD corrections for the fully off-shell leptonic top-decays, without and
with an additional Higgs boson in the final state, i.e., the processes
\begin{equation}
  \label{eq:reaction}
  e^+e^- \to e^+\nu_e\mu^-\bar{\nu}_\mu b \bar{b} (+H)\, ,
\end{equation}
see~\cite{Nejad:2016bci}, as well as the matching of the fully exclusive resummed non-relativistic
threshold corrections within non-relativistic QCD (NRQCD) at
next-to-leading logarithmic accuracy (NLL) to the above mentioned
continuum (NLO QCD) calculation~\cite{Bach:2017ggt}. Both calculations
have been relying on the framework for automated NLO QCD calculations
within the Monte-Carlo event generator
\texttt{WHIZARD}~\cite{Kilian:2007gr} which has been developed during
the final phase of the SFB 676. 

In Ref.~\cite{Nejad:2016bci}, top-pair
production and top-pair production in association with a Higgs boson
have been calculated at three different levels of off-shellness. First,
for the on-shell processes, 
\begin{equation}
  \label{eq:reaction1}
  e^+e^- \to t\bar{t} (H) \, , 
\end{equation}
second, with decaying top-quarks, 
\begin{equation}
  \label{eq:reaction2}
  e^+e^- \to W^- b W^+ \bar{b} (H)\, ,
\end{equation}
and, third, with leptonic $W$-boson decays, 
\begin{equation}
  \label{eq:reaction3}
  e^+e^- \to e^+\nu_e \mu^- \bar{nu}_\mu b \bar{b} (H) \, .
\end{equation}

\begin{figure}[t]
\centering
\includegraphics[width=0.455\textwidth]{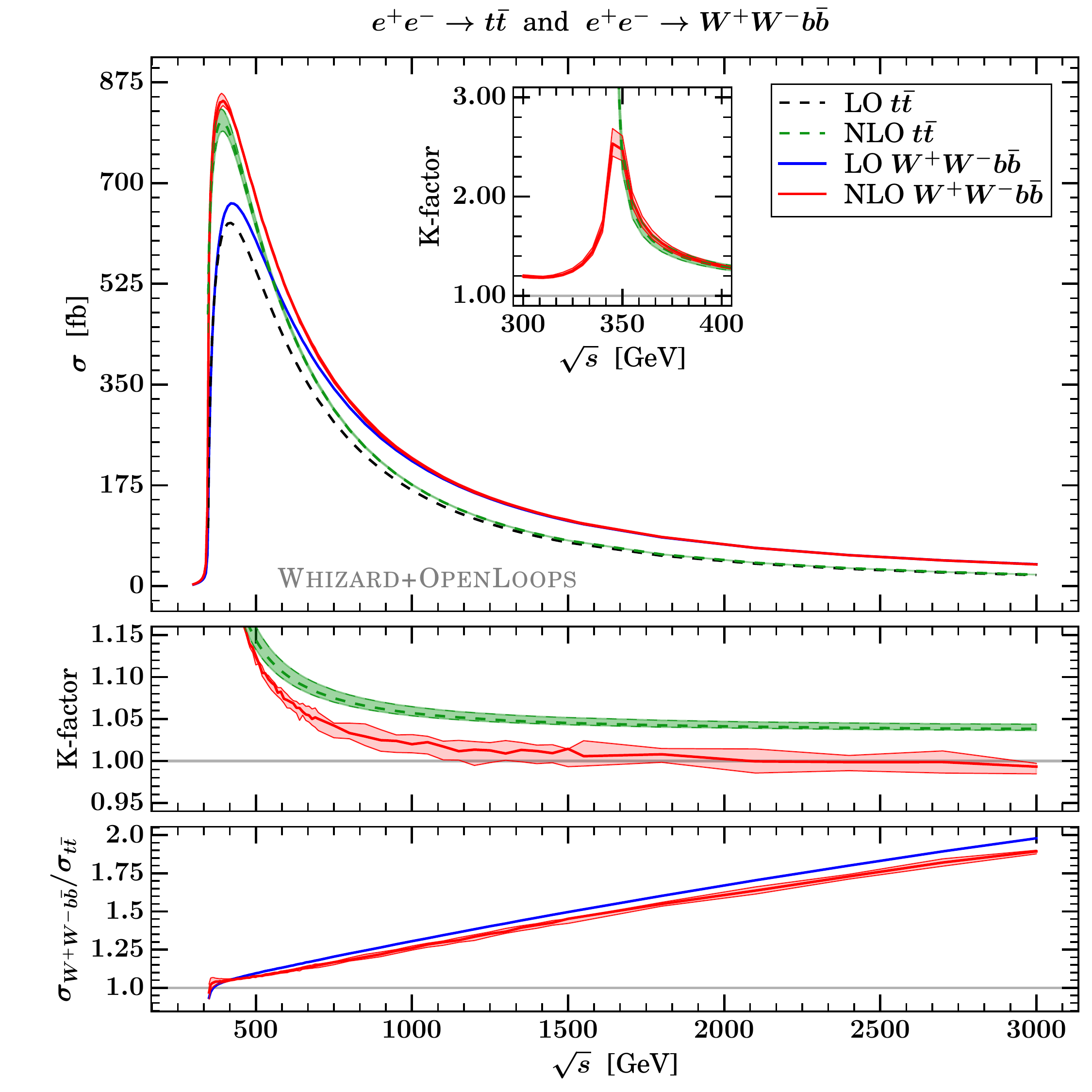}
\hfill
\includegraphics[width=0.53\textwidth]{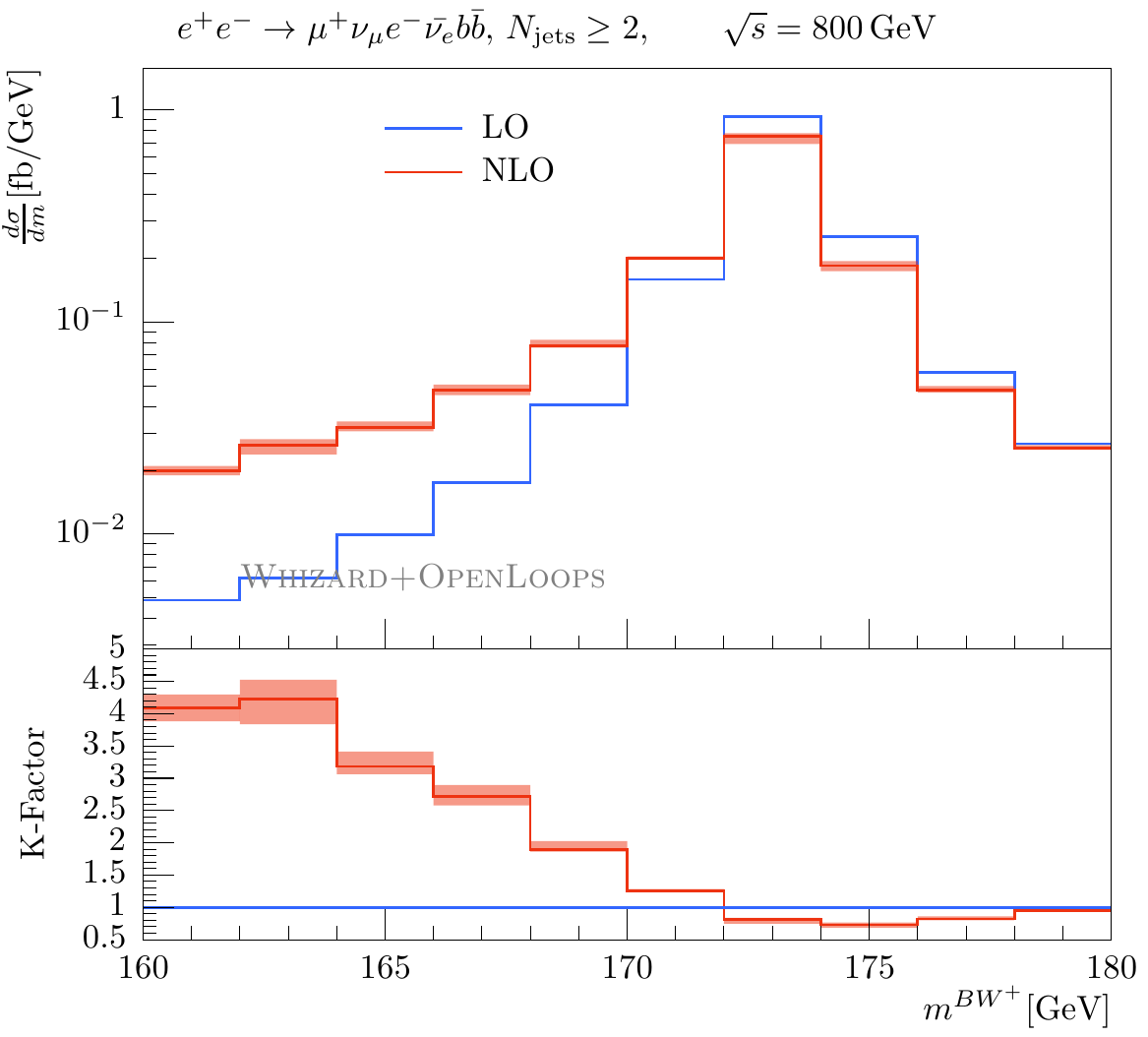}
\caption{Left panel: Total cross section for
  $e^+e^- \to W^+ b W^-\bar{b}$ at leading order (LO) QCD (blue) and NLO QCD
  (red), respectively. Dashed curves show the on-shell process
  $e^+e^-\to t\bar{t}$. The ratio plots show the $K$-factor (NLO QCD
  over LO QCD) and the off-shell over the on-shell process. The inset
  shows the $K$-factor close to threshold. 
  Right panel: Differential
  distribution for the invariant mass of the $b$-jet and
  the $W^+$ for $e^+e^- \to e^+ \nu_e \mu^- \bar{\nu}_\mu b \bar{b}$
  at LO QCD (blue) and NLO QCD (red), respectively. The red band shows
  the scale uncertainties for $m_t/2 < \mu < 2 m_t$ around the
  central scale $\mu = m_t$. 
  Figures taken from Ref.~\cite{Nejad:2016bci}.  
  }
\label{fig:eett}
\end{figure}

\begin{figure}[t]
\centering
\includegraphics[width=0.45\textwidth]{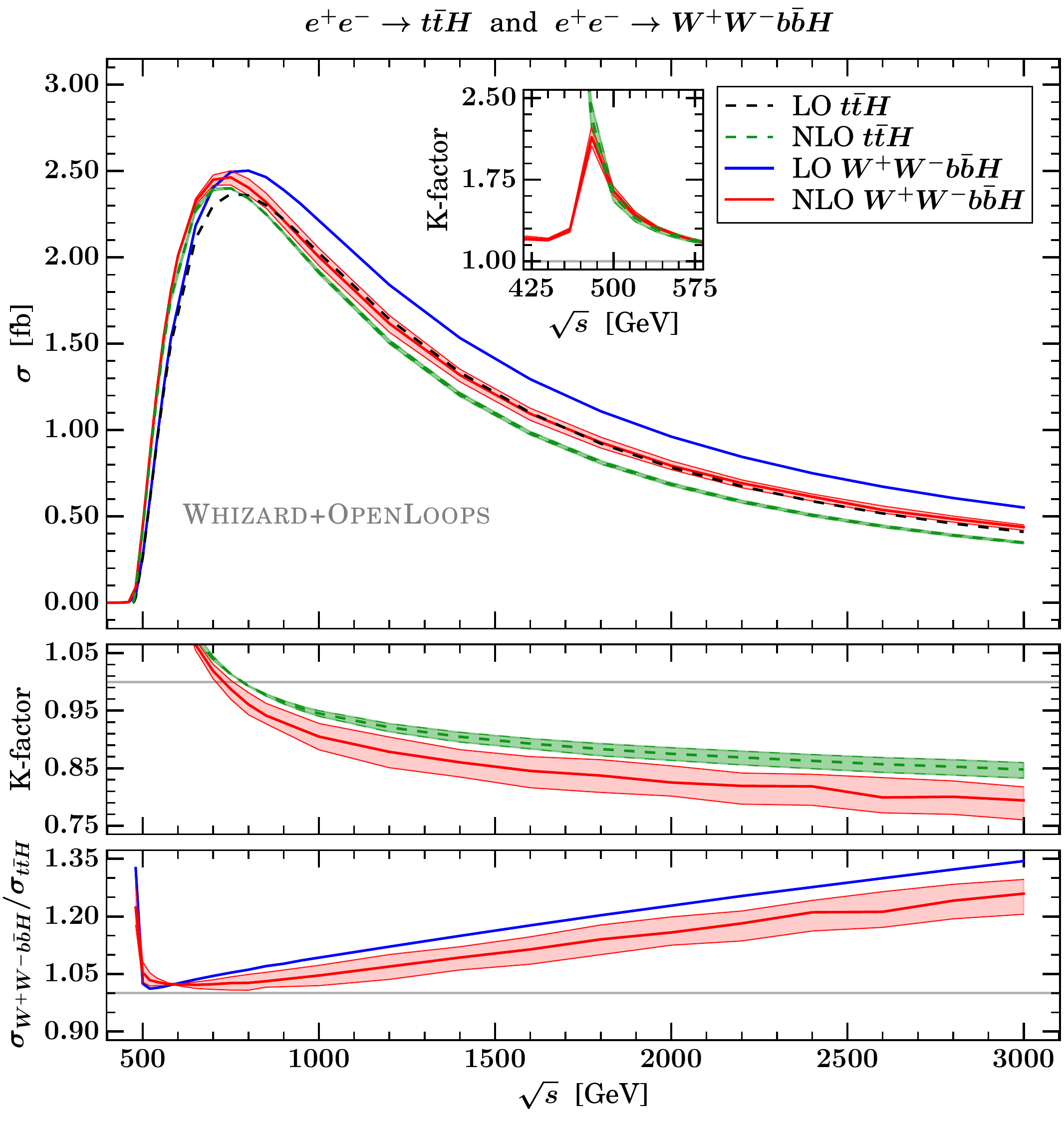}
\hfill
\includegraphics[width=0.54\textwidth]{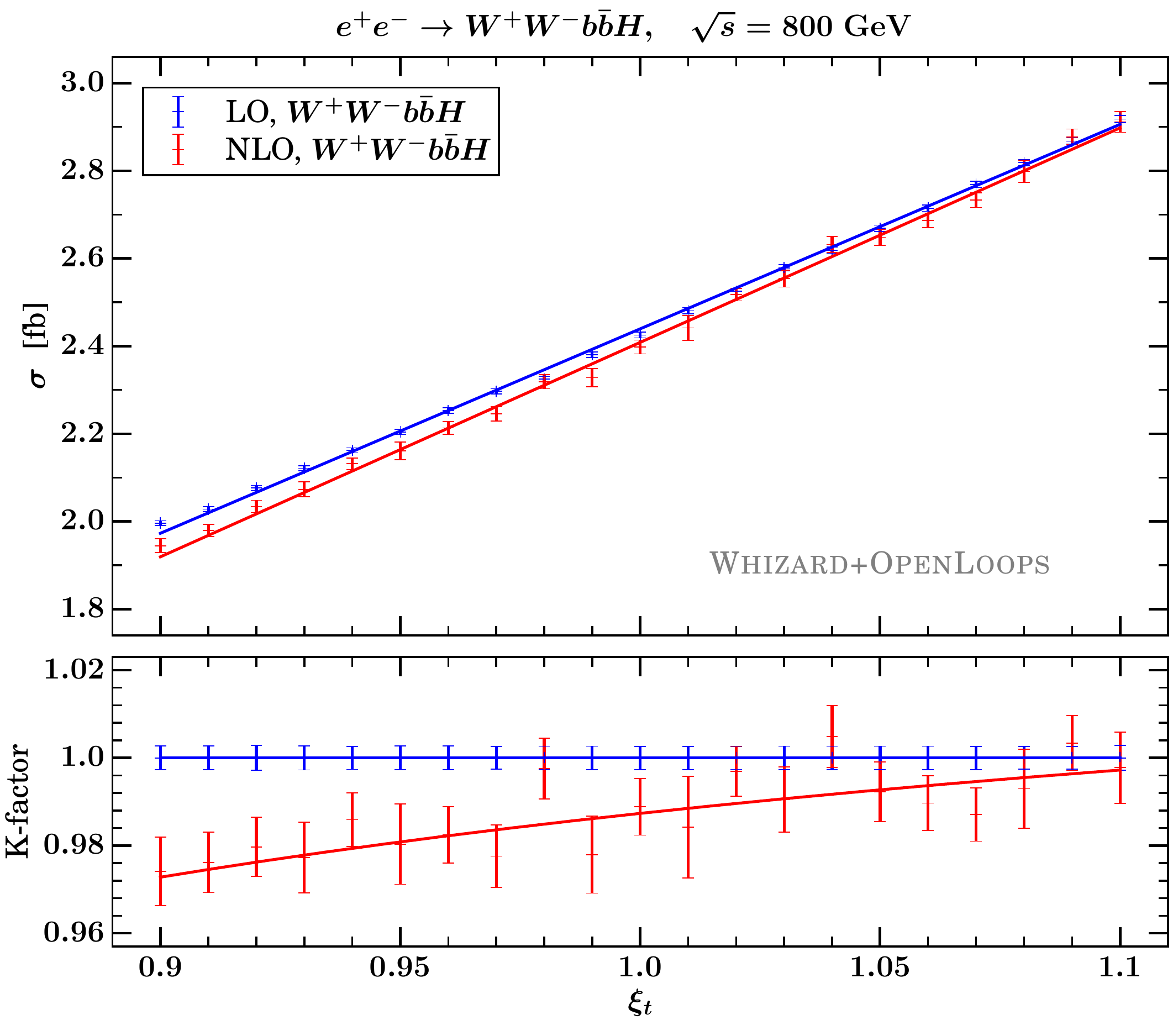}
\caption{Left panel: Total cross section for
  $e^+e^- \to W^+ b W^-\bar{b} H$ at LO QCD (blue) and NLO QCD
  (red), respectively. Dashed curves show the on-shell process
  $e^+e^-\to t\bar{t} H$. The ratio plots show the $K$-factor (NLO QCD
  over LO QCD) and the off-shell over the on-shell process. The inset
  shows the $K$-factor close to threshold. Right panel: Dependence of
  the total cross section for $e^+e^- \to W^+ b W^-\bar{b} H$ at LO QCD
  (blue) and NLO QCD (red), respectively, on the signal strength
  modifier for the top-Yukawa coupling, $\xi_t = y_t /
  y_t^{SM}$. Figures taken from Ref.~\cite{Nejad:2016bci}.
  }
\label{fig:eetth}
\end{figure}

For top-pair production, the main results are shown in Fig.~\ref{fig:eett}. 
The left panel shows the total cross section at
LO and NLO QCD for the full energy range up to the maximal energy of
the CLIC project of $\sqrt{s}=3$~TeV. The two panels at the bottom show the 
$K$-factor (the ratio between NLO and LO total sections), as well as the
ratio of the off-shell to the on-shell process. This ratio is rising
from unity close to threshold up to a factor of two, demonstrating that
the non-resonant irreducible background becomes more and more
important at higher energy. 
The inset shows the $K$-factor in the
proximity of the top-quark threshold, where fixed-order perturbation theory
is not a good approximation any more due to the strong QCD binding
effects of the non-relativistic top-quark pair. The right panel shows
the differential distribution of the invariant mass of the $b$-jet and
the $W^+$ at LO and NLO QCD, with the $K$-factor being clearly not
constant over the phase space. In Fig.~\ref{fig:eetth}, the
corresponding total cross section for off-shell top pair production in
association with a Higgs boson is shown in the left panel. All curves
have the corresponding meaning compared to Fig.~\ref{fig:eett}. The
right panel shows the total cross section for the off-shell process
\begin{equation}
  \label{eq:next-reaction}
  e^+e^- \to W^+ b W^- \bar{b} H \, ,
\end{equation}
as a function of the signal strength
modifier $\xi_t = y_t / y_t^{SM}$ for the top-Yukawa coupling. It
shows that the dependence on this modifier differs between LO and NLO
QCD. The results from this NLO QCD calculation have been used by the
CLICdp collaboration for the assessment on the precision with which
the top-Yukawa coupling can be measured at CLIC for 1.4 TeV
center-of-mass energy~\cite{CLICtop}.  

\begin{figure}[t]
\centering
\includegraphics[width=0.5\textwidth]{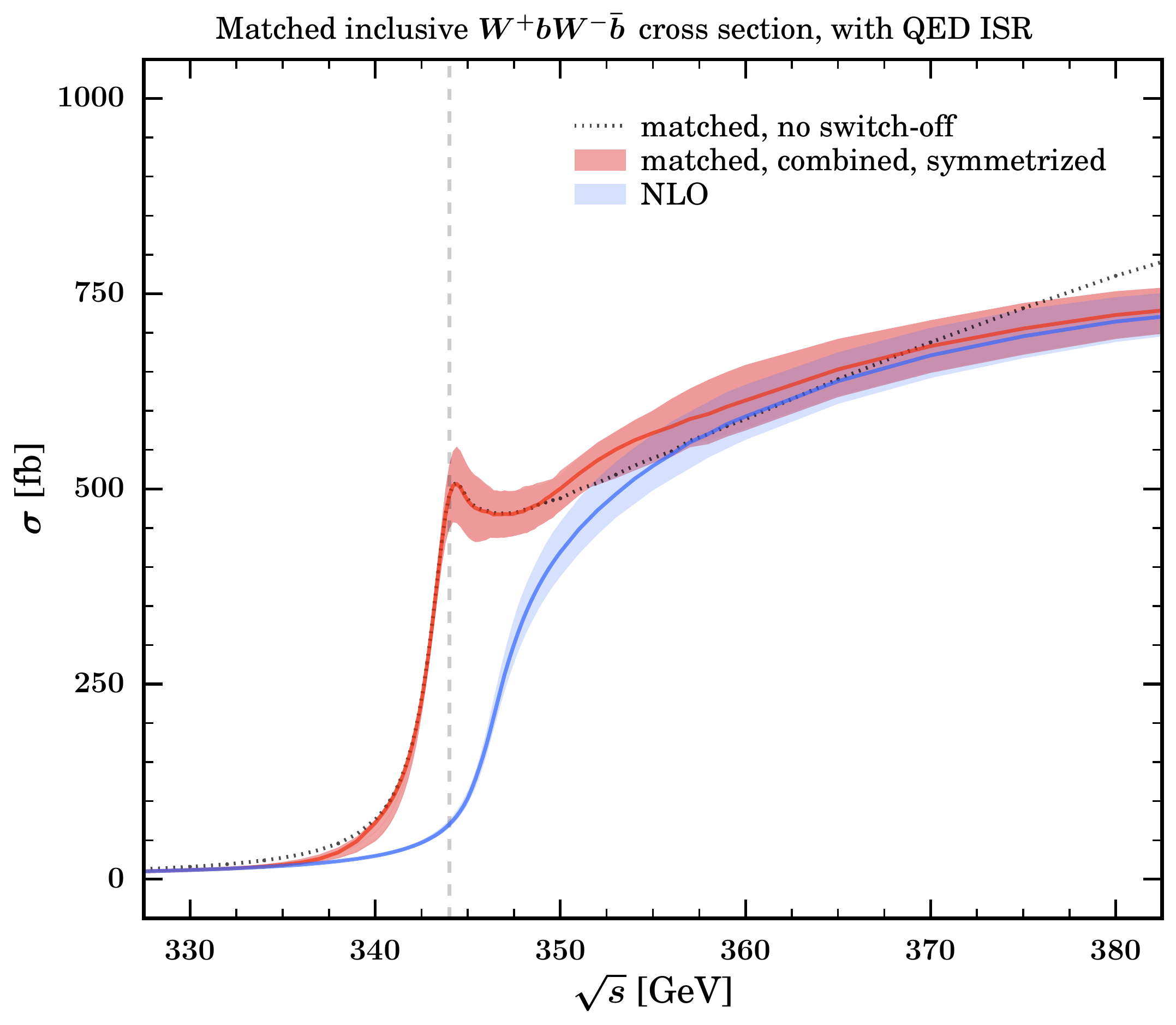}
\hfill
\includegraphics[width=0.49\textwidth]{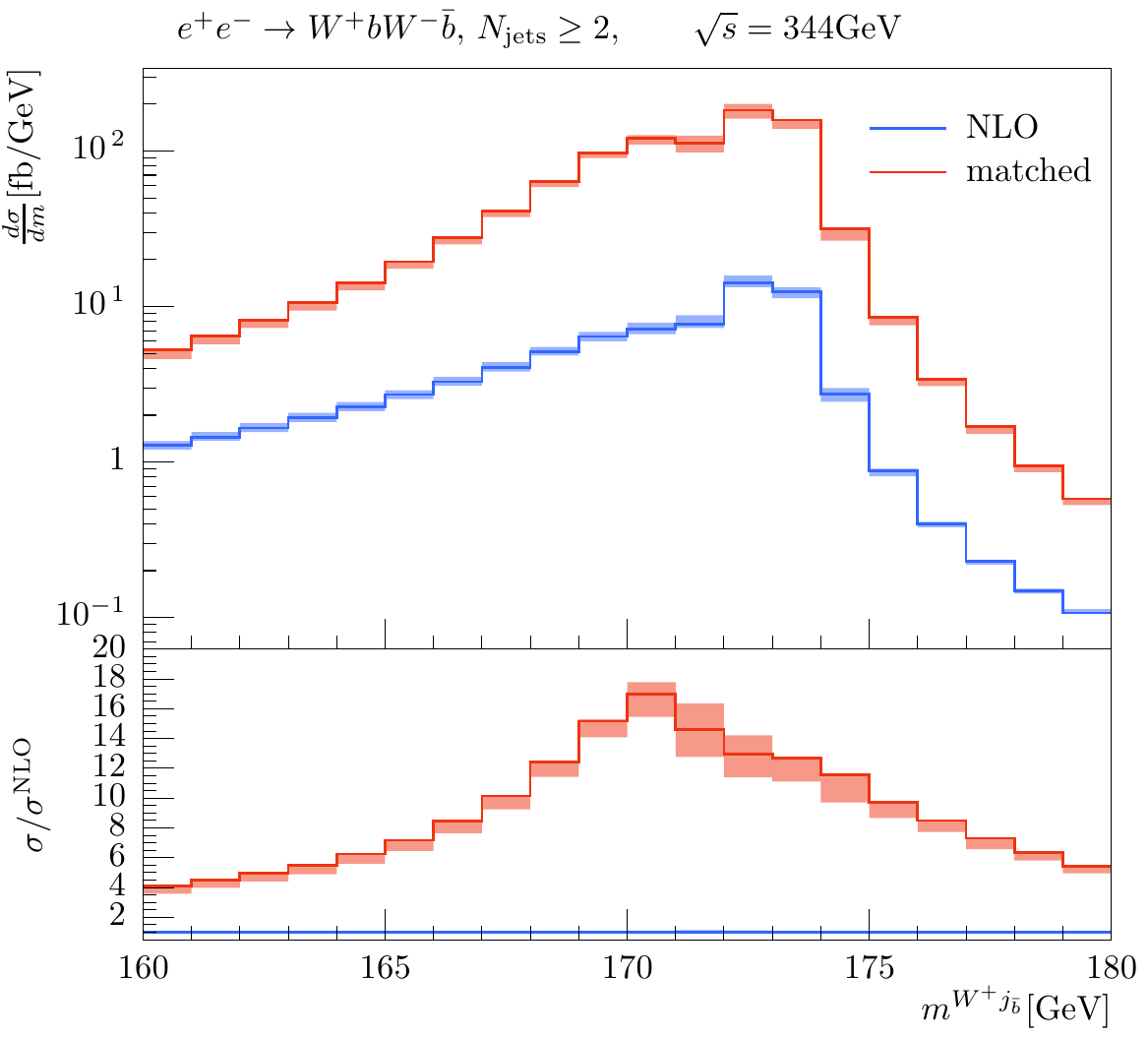}
\caption{Left panel: Threshold scan for the exclusive process $e^+e^-
  \to W^+ b W^- \bar{b}$ matched at NLL NRQCD (threshold) to NLO QCD
  (continuum), including full all-order QED initial state radiation effects. 
  Right panel: Differential distribution for the invariant mass of the $b$-jet and
  the $W+$ boson at fixed order NLO QCD (blue) and fully matched
  (red), respectively. Figures taken from Ref.~\cite{Bach:2017ggt}.  
  }
\label{fig:topthreshold}
\end{figure}

The most precise method known to measure the top-quark mass with an ultimate
precision of $\Delta m_t=30-70$~MeV is a template fit of the top threshold scan for
the total inclusive cross section. The main uncertainties in addition to 
the unknown higher order corrections beyond 
next-to-next-to-next-to-leading order (NNNLO) in QCD are theoretical
background uncertainties, systematic uncertainties due to selection
efficiencies and uncertainties from the beam spectrum. 
All of the last three ones can be addressed with the work
from~\cite{Bach:2017ggt}. The main difficulty is to properly match the
non-relativistic effective-field theory calculation using NRQCD that
is best suited right at the threshold with the relativistic NLO QCD
calculation at the continuum. For that purpose, a factorized
calculation in double-pole approximation with on-shell projection has
been devised in~\cite{Bach:2017ggt} to subtract the double-counting of
$\alpha_s$ corrections from both approaches, thereby maintaining full
electroweak gauge invariance. The left panel of
Fig.~\ref{fig:topthreshold} shows the fully exclusive matched
threshold scan for the process 
\begin{equation}
  \label{eq:next-next-reaction}
  e^+e^- \to W^+ b W^- \bar{b} \, .
\end{equation}
The blue curve shows the fixed-order NLO QCD calculation including scale
variations, while the red curve shows the fully matched
calculation. The red band includes uncertainties from the soft and
hard NRQCD scale variations, the relativistic NLO QCD scale
variations, variations over the details of the switch-off of the
non-relativistic contributions in the continuum. (The dashed black line
shows how the non-relativistic threshold calculation fails to describe
top pair production properly in the continuum.) To be more
conservative, scale variations are also symmetrized with respect to
the central value line as the NRQCD scale variations are highly
asymmetric. The plot includes the effects from an all-order
resummation of soft-collinear QED photons from initial-state radiation
as well as hard-collinear photons up to third order in $\alpha$. The
matched exclusive calculation provided in~\cite{Bach:2017ggt} allows
for the first time to study differential top-quark distributions in
the vicinity of the top threshold. This is shown in the right panel of
Fig.~\ref{fig:topthreshold} for the invariant mass of the $b$-jet and
the $W^+$ boson. The blue curve shows the NLO QCD result, while the
red curve shows the matched calculation.

\section{Conclusions}

Top-quark physics is one the major fields in particle physics, both to
study the properties of the heaviest known particle in Nature, as well
as using it as a means to search for new physics. In this project,
forefront precision calculations have been performed for the studies
of the properties of the top-quark at the running LHC as well as for
future lepton colliders. The calculations made in this project are
very important for the measurement of the mass of the top-quark, the
strong coupling constant, the top-Yukawa coupling and the width of the
top-quark, as well as for a better understanding of the parton
distribution functions of the proton. In addition, top-quark physics
has been used as a tool for searches for new physics, e.g. for the
search of heavy top-like new quarks, so-called top-partner particles.

\subsection*{Acknowledgments}
S.~M. gratefully acknowledges collaborations with S. Alekhin, J. Bl\"umlein, J.~Kieseler, 
K. Lipka, R.~Placakyte and S.~Thier; 
and J.~R.~R. wants to thank for collaborations with A.H.~Hoang, W.~Kilian, J.~Lindert 
and S.~Pozzorini.


\begin{footnotesize}

\bibliographystyle{sfb676}
\bibliography{sfb676_B11}

\providecommand{\href}[2]{#2}\begingroup\raggedright\begin{thebibliography}{10}

\bibitem{Alekhin:2017kpj}
S.~Alekhin, J.~Bl{\"u}mlein, S.~Moch and R.~Placakyte, \emph{{Parton
  distribution functions, $\alpha_s$, and heavy-quark masses for LHC Run II}},
  \href{https://doi.org/10.1103/PhysRevD.96.014011}{\emph{Phys. Rev.}
  {\bfseries D96} (2017) 014011},
  [\href{https://arxiv.org/abs/1701.05838}{{\ttfamily 1701.05838}}].

\bibitem{Baernreuther:2012ws}
P.~B{\"a}rnreuther, M.~Czakon and A.~Mitov, \emph{{Percent Level Precision
  Physics at the Tevatron: First Genuine NNLO QCD Corrections to $q \bar{q} \to
  t \bar{t} + X$}},
  \href{https://doi.org/10.1103/PhysRevLett.109.132001}{\emph{Phys. Rev. Lett.}
  {\bfseries 109} (2012) 132001},
  [\href{https://arxiv.org/abs/1204.5201}{{\ttfamily 1204.5201}}].

\bibitem{Czakon:2012zr}
M.~Czakon and A.~Mitov, \emph{{NNLO corrections to top-pair production at
  hadron colliders: the all-fermionic scattering channels}},
  \href{https://doi.org/10.1007/JHEP12(2012)054}{\emph{JHEP} {\bfseries 12}
  (2012) 054}, [\href{https://arxiv.org/abs/1207.0236}{{\ttfamily 1207.0236}}].

\bibitem{Czakon:2012pz}
M.~Czakon and A.~Mitov, \emph{{NNLO corrections to top pair production at
  hadron colliders: the quark-gluon reaction}},
  \href{https://doi.org/10.1007/JHEP01(2013)080}{\emph{JHEP} {\bfseries 01}
  (2013) 080}, [\href{https://arxiv.org/abs/1210.6832}{{\ttfamily 1210.6832}}].

\bibitem{Czakon:2013goa}
M.~Czakon, P.~Fiedler and A.~Mitov, \emph{{Total Top-Quark Pair-Production
  Cross Section at Hadron Colliders Through $O(\alpha_s^4)$}},
  \href{https://doi.org/10.1103/PhysRevLett.110.252004}{\emph{Phys. Rev. Lett.}
  {\bfseries 110} (2013) 252004},
  [\href{https://arxiv.org/abs/1303.6254}{{\ttfamily 1303.6254}}].

\bibitem{Aliev:2010zk}
M.~Aliev, H.~Lacker, U.~Langenfeld, S.~Moch, P.~Uwer and M.~Wiedermann,
  \emph{{HATHOR: HAdronic Top and Heavy quarks crOss section calculatoR}},
  \href{https://doi.org/10.1016/j.cpc.2010.12.040}{\emph{Comput. Phys. Commun.}
  {\bfseries 182} (2011) 1034--1046},
  [\href{https://arxiv.org/abs/1007.1327}{{\ttfamily 1007.1327}}].

\bibitem{Brucherseifer:2014ama}
M.~Brucherseifer, F.~Caola and K.~Melnikov, \emph{{On the NNLO QCD corrections
  to single-top production at the LHC}},
  \href{https://doi.org/10.1016/j.physletb.2014.06.075}{\emph{Phys. Lett.}
  {\bfseries B736} (2014) 58--63},
  [\href{https://arxiv.org/abs/1404.7116}{{\ttfamily 1404.7116}}].

\bibitem{Berger:2016oht}
E.~L. Berger, J.~Gao, C.~P. Yuan and H.~X. Zhu, \emph{{NNLO QCD Corrections to
  t-channel Single Top-Quark Production and Decay}},
  \href{https://doi.org/10.1103/PhysRevD.94.071501}{\emph{Phys. Rev.}
  {\bfseries D94} (2016) 071501},
  [\href{https://arxiv.org/abs/1606.08463}{{\ttfamily 1606.08463}}].

\bibitem{Alekhin:2015cza}
S.~Alekhin, J.~Bl{\"u}mlein, S.~Moch and R.~Placakyte, \emph{{Isospin asymmetry
  of quark distributions and implications for single top-quark production at
  the LHC}}, \href{https://doi.org/10.1103/PhysRevD.94.114038}{\emph{Phys.
  Rev.} {\bfseries D94} (2016) 114038},
  [\href{https://arxiv.org/abs/1508.07923}{{\ttfamily 1508.07923}}].

\bibitem{Kant:2014oha}
P.~Kant, O.~M. Kind, T.~Kintscher, T.~Lohse, T.~Martini, S.~M{\"o}lbitz et~al.,
  \emph{{HatHor for single top-quark production: Updated predictions and
  uncertainty estimates for single top-quark production in hadronic
  collisions}}, \href{https://doi.org/10.1016/j.cpc.2015.02.001}{\emph{Comput.
  Phys. Commun.} {\bfseries 191} (2015) 74--89},
  [\href{https://arxiv.org/abs/1406.4403}{{\ttfamily 1406.4403}}].

\bibitem{Aad:2014fwa}
{\scshape ATLAS} Collaboration, G.~Aad et~al., \emph{{Comprehensive
  measurements of $t$-channel single top-quark production cross sections at
  $\sqrt{s} = 7$ TeV with the ATLAS detector}},
  \href{https://doi.org/10.1103/PhysRevD.90.112006}{\emph{Phys. Rev.}
  {\bfseries D90} (2014) 112006},
  [\href{https://arxiv.org/abs/1406.7844}{{\ttfamily 1406.7844}}].

\bibitem{Khachatryan:2014iya}
{\scshape CMS} Collaboration, V.~Khachatryan et~al., \emph{{Measurement of the
  t-channel single-top-quark production cross section and of the $\mid V_{tb}
  \mid$ CKM matrix element in pp collisions at $\sqrt{s}$= 8 TeV}},
  \href{https://doi.org/10.1007/JHEP06(2014)090}{\emph{JHEP} {\bfseries 06}
  (2014) 090}, [\href{https://arxiv.org/abs/1403.7366}{{\ttfamily 1403.7366}}].

\bibitem{Dulat:2015mca}
S.~Dulat, T.-J. Hou, J.~Gao, M.~Guzzi, J.~Huston, P.~Nadolsky et~al.,
  \emph{{New parton distribution functions from a global analysis of quantum
  chromodynamics}},
  \href{https://doi.org/10.1103/PhysRevD.93.033006}{\emph{Phys. Rev.}
  {\bfseries D93} (2016) 033006},
  [\href{https://arxiv.org/abs/1506.07443}{{\ttfamily 1506.07443}}].

\bibitem{Harland-Lang:2014zoa}
L.~A. Harland-Lang, A.~D. Martin, P.~Motylinski and R.~S. Thorne, \emph{{Parton
  distributions in the LHC era: MMHT 2014 PDFs}},
  \href{https://doi.org/10.1140/epjc/s10052-015-3397-6}{\emph{Eur. Phys. J.}
  {\bfseries C75} (2015) 204},
  [\href{https://arxiv.org/abs/1412.3989}{{\ttfamily 1412.3989}}].

\bibitem{Ball:2017nwa}
{\scshape NNPDF} Collaboration, R.~D. Ball et~al., \emph{{Parton distributions
  from high-precision collider data}},
  \href{https://doi.org/10.1140/epjc/s10052-017-5199-5}{\emph{Eur. Phys. J.}
  {\bfseries C77} (2017) 663},
  [\href{https://arxiv.org/abs/1706.00428}{{\ttfamily 1706.00428}}].

\bibitem{Alekhin:2016jjz}
S.~Alekhin, S.~Moch and S.~Thier, \emph{{Determination of the top-quark mass
  from hadro-production of single top-quarks}},
  \href{https://doi.org/10.1016/j.physletb.2016.10.062}{\emph{Phys. Lett.}
  {\bfseries B763} (2016) 341--346},
  [\href{https://arxiv.org/abs/1608.05212}{{\ttfamily 1608.05212}}].

\bibitem{Moch:2015fra}
S.~Moch, \emph{{Precise heavy-quark masses}},
  \href{https://doi.org/10.1016/j.nuclphysbps.2015.03.011}{\emph{Nucl. Part.
  Phys. Proc.} {\bfseries 261-262} (2015) 130--139}.

\bibitem{Marquard:2016dcn}
P.~Marquard, A.~V. Smirnov, V.~A. Smirnov, M.~Steinhauser and D.~Wellmann,
  \emph{{$\overline{\rm MS}$-on-shell quark mass relation up to four loops in
  QCD and a general SU$(N)$ gauge group}},
  \href{https://doi.org/10.1103/PhysRevD.94.074025}{\emph{Phys. Rev.}
  {\bfseries D94} (2016) 074025},
  [\href{https://arxiv.org/abs/1606.06754}{{\ttfamily 1606.06754}}].

\bibitem{Moch:2014tta}
S.~Moch et~al., \emph{{High precision fundamental constants at the TeV scale}},
   \href{https://arxiv.org/abs/1405.4781}{{\ttfamily 1405.4781}}.

\bibitem{Langenfeld:2009wd}
U.~Langenfeld, S.~Moch and P.~Uwer, \emph{{Measuring the running top-quark
  mass}}, \href{https://doi.org/10.1103/PhysRevD.80.054009}{\emph{Phys. Rev.}
  {\bfseries D80} (2009) 054009},
  [\href{https://arxiv.org/abs/0906.5273}{{\ttfamily 0906.5273}}].

\bibitem{Kieseler:2015jzh}
J.~Kieseler, K.~Lipka and S.-O. Moch, \emph{{Calibration of the Top-Quark Monte
  Carlo Mass}},
  \href{https://doi.org/10.1103/PhysRevLett.116.162001}{\emph{Phys. Rev. Lett.}
  {\bfseries 116} (2016) 162001},
  [\href{https://arxiv.org/abs/1511.00841}{{\ttfamily 1511.00841}}].

\bibitem{Aad:2015zhl}
{\scshape ATLAS, CMS} Collaborations, G.~Aad et~al., \emph{{Combined
  Measurement of the Higgs Boson Mass in $pp$ Collisions at $\sqrt{s}=7$ and 8
  TeV with the ATLAS and CMS Experiments}},
  \href{https://doi.org/10.1103/PhysRevLett.114.191803}{\emph{Phys. Rev. Lett.}
  {\bfseries 114} (2015) 191803},
  [\href{https://arxiv.org/abs/1503.07589}{{\ttfamily 1503.07589}}].

\bibitem{Buttazzo:2013uya}
D.~Buttazzo, G.~Degrassi, P.~P. Giardino, G.~F. Giudice, F.~Sala, A.~Salvio
  et~al., \emph{{Investigating the near-criticality of the Higgs boson}},
  \href{https://doi.org/10.1007/JHEP12(2013)089}{\emph{JHEP} {\bfseries 12}
  (2013) 089}, [\href{https://arxiv.org/abs/1307.3536}{{\ttfamily 1307.3536}}].

\bibitem{Bednyakov:2015sca}
A.~V. Bednyakov, B.~A. Kniehl, A.~F. Pikelner and O.~L. Veretin,
  \emph{{Stability of the Electroweak Vacuum: Gauge Independence and Advanced
  Precision}},
  \href{https://doi.org/10.1103/PhysRevLett.115.201802}{\emph{Phys. Rev. Lett.}
  {\bfseries 115} (2015) 201802},
  [\href{https://arxiv.org/abs/1507.08833}{{\ttfamily 1507.08833}}].

\bibitem{Alekhin:2012py}
S.~Alekhin, A.~Djouadi and S.~Moch, \emph{{The top quark and Higgs boson masses
  and the stability of the electroweak vacuum}},
  \href{https://doi.org/10.1016/j.physletb.2012.08.024}{\emph{Phys. Lett.}
  {\bfseries B716} (2012) 214--219},
  [\href{https://arxiv.org/abs/1207.0980}{{\ttfamily 1207.0980}}].

\bibitem{Kniehl:2016enc}
B.~A. Kniehl, A.~F. Pikelner and O.~L. Veretin, \emph{{mr: a C++ library for
  the matching and running of the Standard Model parameters}},
  \href{https://doi.org/10.1016/j.cpc.2016.04.017}{\emph{Comput. Phys. Commun.}
  {\bfseries 206} (2016) 84--96},
  [\href{https://arxiv.org/abs/1601.08143}{{\ttfamily 1601.08143}}].

\bibitem{Reuter:2014iya}
J.~Reuter and M.~Tonini, \emph{{Top Partner Discovery in the T $\to$ tZ channel
  at the LHC}}, \href{https://doi.org/10.1007/JHEP01(2015)088}{\emph{JHEP}
  {\bfseries 01} (2015) 088},
  [\href{https://arxiv.org/abs/1409.6962}{{\ttfamily 1409.6962}}].

\bibitem{Reuter:2012sd}
J.~Reuter and M.~Tonini, \emph{{Can the 125 GeV Higgs be the Little Higgs?}},
  \href{https://doi.org/10.1007/JHEP02(2013)077}{\emph{JHEP} {\bfseries 02}
  (2013) 077}, [\href{https://arxiv.org/abs/1212.5930}{{\ttfamily 1212.5930}}].

\bibitem{Reuter:2013iya}
J.~Reuter, M.~Tonini and M.~de~Vries, \emph{{Littlest Higgs with T-parity:
  Status and Prospects}},
  \href{https://doi.org/10.1007/JHEP02(2014)053}{\emph{JHEP} {\bfseries 02}
  (2014) 053}, [\href{https://arxiv.org/abs/1310.2918}{{\ttfamily 1310.2918}}].

\bibitem{Dercks:2018hgz}
D.~Dercks, G.~Moortgat-Pick, J.~Reuter and S.~Y. Shim, \emph{{The fate of the
  Littlest Higgs Model with T-parity under 13 TeV LHC Data}},
  \href{https://doi.org/10.1007/JHEP05(2018)049}{\emph{JHEP} {\bfseries 05}
  (2018) 049}, [\href{https://arxiv.org/abs/1801.06499}{{\ttfamily
  1801.06499}}].

\bibitem{Bach:2014zca}
F.~Bach and T.~Ohl, \emph{{Anomalous top charged-current contact interactions
  in single top production at the LHC}},
  \href{https://doi.org/10.1103/PhysRevD.90.074022}{\emph{Phys. Rev.}
  {\bfseries D90} (2014) 074022},
  [\href{https://arxiv.org/abs/1407.2546}{{\ttfamily 1407.2546}}].

\bibitem{Baer:2013cma}
H.~Baer, T.~Barklow, K.~Fujii, Y.~Gao, A.~Hoang, S.~Kanemura et~al., \emph{{The
  International Linear Collider Technical Design Report - Volume 2: Physics}},
  \href{https://arxiv.org/abs/1306.6352}{{\ttfamily 1306.6352}}.

\bibitem{Fujii:2015jha}
K.~Fujii et~al., \emph{{Physics Case for the International Linear Collider}},
  \href{https://arxiv.org/abs/1506.05992}{{\ttfamily 1506.05992}}.

\bibitem{Linssen:2012hp}
L.~Linssen, A.~Miyamoto, M.~Stanitzki and H.~Weerts, \emph{{Physics and
  Detectors at CLIC: CLIC Conceptual Design Report}},
  \href{https://arxiv.org/abs/1202.5940}{{\ttfamily 1202.5940}}.

\bibitem{Lebrun:2012hj}
P.~Lebrun, L.~Linssen, A.~Lucaci-Timoce, D.~Schulte, F.~Simon, S.~Stapnes
  et~al., \emph{{The CLIC Programme: Towards a Staged e+e- Linear Collider
  Exploring the Terascale : CLIC Conceptual Design Report}},
  \href{https://arxiv.org/abs/1209.2543}{{\ttfamily 1209.2543}}.

\bibitem{Barklow:2015tja}
T.~Barklow, J.~Brau, K.~Fujii, J.~Gao, J.~List, N.~Walker et~al., \emph{{ILC
  Operating Scenarios}},  \href{https://arxiv.org/abs/1506.07830}{{\ttfamily
  1506.07830}}.

\bibitem{Nejad:2016bci}
B.~Chokouf{\'e}~Nejad, W.~Kilian, J.~M. Lindert, S.~Pozzorini, J.~Reuter and
  C.~Weiss, \emph{{NLO QCD predictions for off-shell $ t\overline{t} $ and $
  t\overline{t}H $ production and decay at a linear collider}},
  \href{https://doi.org/10.1007/JHEP12(2016)075}{\emph{JHEP} {\bfseries 12}
  (2016) 075}, [\href{https://arxiv.org/abs/1609.03390}{{\ttfamily
  1609.03390}}].

\bibitem{Bach:2017ggt}
F.~Bach, B.~C. Nejad, A.~Hoang, W.~Kilian, J.~Reuter, M.~Stahlhofen et~al.,
  \emph{{Fully-differential Top-Pair Production at a Lepton Collider: From
  Threshold to Continuum}},
  \href{https://doi.org/10.1007/JHEP03(2018)184}{\emph{JHEP} {\bfseries 03}
  (2018) 184}, [\href{https://arxiv.org/abs/1712.02220}{{\ttfamily
  1712.02220}}].

\bibitem{Kilian:2007gr}
W.~Kilian, T.~Ohl and J.~Reuter, \emph{{WHIZARD: Simulating Multi-Particle
  Processes at LHC and ILC}},
  \href{https://doi.org/10.1140/epjc/s10052-011-1742-y}{\emph{Eur. Phys. J.}
  {\bfseries C71} (2011) 1742},
  [\href{https://arxiv.org/abs/0708.4233}{{\ttfamily 0708.4233}}].

\bibitem{CLICtop}
{\scshape CLICdp} Collaboration, H.~Abramowicz et~al., \emph{{Top-Quark Physics
  at the CLIC Electron-Positron Linear Collider}},
  \href{https://arxiv.org/abs/1807.02441}{{\ttfamily 1807.02441}}.

\end{thebibliography}\endgroup

\end{footnotesize}


\end{document}